\pdfoutput=1
\documentclass[12pt,a4paper]{article}

\usepackage{comment}
\usepackage{ifthen} 
\newboolean{pdflatex}
\setboolean{pdflatex}{true} 

\newboolean{articletitles}
\setboolean{articletitles}{true} 

\newboolean{uprightparticles}
\setboolean{uprightparticles}{false} 

\def\paperauthors{LHCb collaboration} 
\def\paperasciititle{Test of lepton universality in b->sl^+l^- decays} 
\def\papertitle{Test of lepton universality in \bsll  decays} 
\def\paperkeywords{{High Energy Physics}, {LHCb}} 
\def\papercopyright{\the\year\ CERN for the benefit of the LHCb collaboration} 
\def\paperlicence{CC BY 4.0 licence}
\def\paperlicenceurl{https://creativecommons.org/licenses/by/4.0/}

\usepackage[top=1in, bottom=1.25in, left=1in, right=1in]{geometry}

\usepackage{microtype}
\usepackage{lineno}  
\usepackage{xspace} 
\usepackage{caption} 

\usepackage{graphicx}  
\usepackage{color}
\usepackage{colortbl}
\graphicspath{{./figs/}} 

\usepackage{amsmath} 
\usepackage{amssymb}
\usepackage{amsfonts}
\usepackage{upgreek} 

\newcommand*\patchAmsMathEnvironmentForLineno[1]{%
\expandafter\let\csname old#1\expandafter\endcsname\csname #1\endcsname
\expandafter\let\csname oldend#1\expandafter\endcsname\csname
end#1\endcsname
 \renewenvironment{#1}%
   {\linenomath\csname old#1\endcsname}%
   {\csname oldend#1\endcsname\endlinenomath}%
}
\newcommand*\patchBothAmsMathEnvironmentsForLineno[1]{%
  \patchAmsMathEnvironmentForLineno{#1}%
  \patchAmsMathEnvironmentForLineno{#1*}%
}
\AtBeginDocument{%
\patchBothAmsMathEnvironmentsForLineno{equation}%
\patchBothAmsMathEnvironmentsForLineno{align}%
\patchBothAmsMathEnvironmentsForLineno{flalign}%
\patchBothAmsMathEnvironmentsForLineno{alignat}%
\patchBothAmsMathEnvironmentsForLineno{gather}%
\patchBothAmsMathEnvironmentsForLineno{multline}%
\patchBothAmsMathEnvironmentsForLineno{eqnarray}%
}

\usepackage{hyperxmp}

\usepackage[pdftex,
            pdfauthor={\paperauthors},
            pdftitle={\paperasciititle},
            pdfkeywords={\paperkeywords},
            pdfcopyright={Copyright (C) \papercopyright},
            pdflicenseurl={\paperlicenceurl}]{hyperref}

\usepackage[colorinlistoftodos,textsize=scriptsize]{todonotes}

\usepackage[bottom,flushmargin,hang,multiple]{footmisc}

\usepackage[all]{hypcap} 

\usepackage{xspace} 
\usepackage{upgreek}

\def\lhcb   {\mbox{LHCb}\xspace}

\def\lhc    {\mbox{LHC}\xspace}

\def\rich   {RICH\xspace}

\def\MagUp {\mbox{\em Mag\kern -0.05em Up}\xspace}

\def\lone   {L0\xspace}
\def\hlt    {HLT\xspace}

\ifthenelse{\boolean{uprightparticles}}%
{

 \def\Pmu         {\ensuremath{\upmu}\xspace}                 
 \def\Pnu         {\ensuremath{\upnu}\xspace}                 
                  
 \def\Ppi         {\ensuremath{\uppi}\xspace}

 \def\Ppsi        {\ensuremath{\uppsi}\xspace}

 \def\PDelta      {\ensuremath{\Delta}\xspace}                 
 \def\PXi         {\ensuremath{\Xi}\xspace}                 
 \def\PLambda     {\ensuremath{\Lambda}\xspace}                 
 \def\PSigma      {\ensuremath{\Sigma}\xspace}                 
 \def\POmega      {\ensuremath{\Omega}\xspace}                 
 \def\PUpsilon    {\ensuremath{\Upsilon}\xspace}
 \let\oldPi\Pi
 \def\PPi         {\ensuremath{\oldPi}\xspace}

 \def\PB      {\ensuremath{\mathrm{B}}\xspace}                 
                  
 \def\PD      {\ensuremath{\mathrm{D}}\xspace}

 \def\PJ      {\ensuremath{\mathrm{J}}\xspace}                 
 \def\PK      {\ensuremath{\mathrm{K}}\xspace}

 \def\PW      {\ensuremath{\mathrm{W}}\xspace}

 \def\PZ      {\ensuremath{\mathrm{Z}}\xspace}                 
                  
 \def\Pb      {\ensuremath{\mathrm{b}}\xspace}

 \def\Pe      {\ensuremath{\mathrm{e}}\xspace}

 \def\Pi      {\ensuremath{\mathrm{i}}\xspace}

 \def\Pp      {\ensuremath{\mathrm{p}}\xspace}

 \def\Ps      {\ensuremath{\mathrm{s}}\xspace}

 \def\thebaroffset{0.0em}
}
{

 \def\Pmu         {\ensuremath{\mu}\xspace}                 
 \def\Pnu         {\ensuremath{\nu}\xspace}                 
                  
 \def\Ppi         {\ensuremath{\pi}\xspace}

 \def\Ppsi        {\ensuremath{\psi}\xspace}                 
                  
 \mathchardef\PDelta="7101
 \mathchardef\PXi="7104
 \mathchardef\PLambda="7103
 \mathchardef\PSigma="7106
 \mathchardef\POmega="710A
 \mathchardef\PUpsilon="7107
 \mathchardef\PPi="7105
                  
 \def\PB      {\ensuremath{B}\xspace}                 
                  
 \def\PD      {\ensuremath{D}\xspace}

 \def\PJ      {\ensuremath{J}\xspace}                 
 \def\PK      {\ensuremath{K}\xspace}

 \def\PW      {\ensuremath{W}\xspace}

 \def\PZ      {\ensuremath{Z}\xspace}                 
                  
 \def\Pb      {\ensuremath{b}\xspace}

 \def\Pe      {\ensuremath{e}\xspace}

 \def\Pi      {\ensuremath{i}\xspace}

 \def\Pp      {\ensuremath{p}\xspace}

 \def\Ps      {\ensuremath{s}\xspace}

 \def\thebaroffset{0.18em}
}
\newcommand{\offsetoverline}[2][\thebaroffset]{\kern #1\overline{\kern -#1 #2}}%

\makeatletter
\ifcase \@ptsize \relax
  \newcommand{\miniscule}{\@setfontsize\miniscule{4}{5}}
\or
  \newcommand{\miniscule}{\@setfontsize\miniscule{5}{6}}
\or
  \newcommand{\miniscule}{\@setfontsize\miniscule{5}{6}}
\fi
\makeatother

\DeclareRobustCommand{\optbar}[1]{\shortstack{{\miniscule (\rule[.5ex]{1.25em}{.18mm})}
  \\ [-.7ex] $#1$}}

\def\electron   {{\ensuremath{\Pe}}\xspace}

\def\epem       {{\ensuremath{\Pe^+\Pe^-}}\xspace}

\def\mumu       {{\ensuremath{\Pmu^+\Pmu^-}}\xspace}

\def\ellm       {{\ensuremath{\ell^-}}\xspace}
\def\ellp       {{\ensuremath{\ell^+}}\xspace}

\def\ellell     {\ensuremath{\ell^+ \ell^-}\xspace}

\def\neu        {{\ensuremath{\Pnu}}\xspace}

\def\neue       {{\ensuremath{\neu_e}}\xspace}

\def\W      {{\ensuremath{\PW}}\xspace}

\def\Z      {{\ensuremath{\PZ}}\xspace}

\def\squark    {{\ensuremath{\Ps}}\xspace}

\def\bquark    {{\ensuremath{\Pb}}\xspace}

\def\pion   {{\ensuremath{\Ppi}}\xspace}

\def\pim    {{\ensuremath{\pion^-}}\xspace}

\def\kaon    {{\ensuremath{\PK}}\xspace}

\def\KorKbar {\kern \thebaroffset\optbar{\kern -\thebaroffset \PK}{}\xspace}

\def\Kp      {{\ensuremath{\kaon^+}}\xspace}

\def\Kstarz  {{\ensuremath{\kaon^{*0}}}\xspace}

\def\Kstar   {{\ensuremath{\kaon^*}}\xspace}

\def\D       {{\ensuremath{\PD}}\xspace}

\def\DorDbar {\kern \thebaroffset\optbar{\kern -\thebaroffset \PD}\xspace}

\def\Dp      {{\ensuremath{\D^+}}\xspace}
\def\Dm      {{\ensuremath{\D^-}}\xspace}

\def\DpDm    {\ensuremath{\Dp {\kern -0.16em \Dm}}\xspace}

\def\B       {{\ensuremath{\PB}}\xspace}

\def\BorBbar {\kern \thebaroffset\optbar{\kern -\thebaroffset \PB}\xspace}

\def\Bd      {{\ensuremath{\B^0}}\xspace}

\def\BdorBdbar {\kern \thebaroffset\optbar{\kern -\thebaroffset \Bd}\xspace}
\def\Bu      {{\ensuremath{\B^+}}\xspace}

\def\Bs      {{\ensuremath{\B^0_\squark}}\xspace}

\def\BsorBsbar {\kern \thebaroffset\optbar{\kern -\thebaroffset \Bs}\xspace}

\def\jpsi     {{\ensuremath{{\PJ\mskip -3mu/\mskip -2mu\Ppsi}}}\xspace}
\def\psitwos  {{\ensuremath{\Ppsi{(2S)}}}\xspace}

\def\Y#1S{\ensuremath{\PUpsilon{(#1S)}}\xspace}

\def\proton      {{\ensuremath{\Pp}}\xspace}

\def\LorLbar     {\kern \thebaroffset\optbar{\kern -\thebaroffset \PLambda}\xspace}

\newcommand{\decay}[2]{\ensuremath{#1\!\to #2}\xspace} 

\def\to                 {\ensuremath{\rightarrow}\xspace}

\def\qsq       {{\ensuremath{q^2}}\xspace}

\def\bsll     {\decay{\bquark}{\squark \ell^+ \ell^-}}

\def\AT#1     {\ensuremath{A_{\mathrm{T}}^{#1}}\xspace}           

\def\C#1      {\ensuremath{\mathcal{C}_{#1}}\xspace}                       
\def\Cp#1     {\ensuremath{\mathcal{C}_{#1}^{'}}\xspace}                    
\def\Ceff#1   {\ensuremath{\mathcal{C}_{#1}^{\mathrm{(eff)}}}\xspace}        
\def\Cpeff#1  {\ensuremath{\mathcal{C}_{#1}^{'\mathrm{(eff)}}}\xspace}       
\def\Ope#1    {\ensuremath{\mathcal{O}_{#1}}\xspace}                       
\def\Opep#1   {\ensuremath{\mathcal{O}_{#1}^{'}}\xspace}                    

\newcommand{\aunit}[1]{\ensuremath{\text{\,#1}}}       

\newcommand{\tev}{\aunit{Te\kern -0.1em V}\xspace}
\newcommand{\gev}{\aunit{Ge\kern -0.1em V}\xspace}
\newcommand{\mev}{\aunit{Me\kern -0.1em V}\xspace}
\newcommand{\kev}{\aunit{ke\kern -0.1em V}\xspace}
\newcommand{\ev}{\aunit{e\kern -0.1em V}\xspace}
 
\newcommand{\mevc}{\ensuremath{\aunit{Me\kern -0.1em V\!/}c}\xspace}
\newcommand{\gevc}{\ensuremath{\aunit{Ge\kern -0.1em V\!/}c}\xspace}
\newcommand{\mevcc}{\ensuremath{\aunit{Me\kern -0.1em V\!/}c^2}\xspace}
\newcommand{\gevcc}{\ensuremath{\aunit{Ge\kern -0.1em V\!/}c^2}\xspace}
\newcommand{\gevgevcccc}{\ensuremath{\gev^2\!/c^4}\xspace} 

\def\fb   {\ensuremath{\aunit{fb}}\xspace}
\def\invfb   {\ensuremath{\fb^{-1}}\xspace}

\newcommand{\stat}{\aunit{(stat)}\xspace}
\newcommand{\syst}{\aunit{(syst)}\xspace}

\newcommand{\chisq}{\ensuremath{\chi^2}\xspace}

\newcommand{\chisqip}{\ensuremath{\chi^2_{\text{IP}}}\xspace}

\def\deriv {\ensuremath{\mathrm{d}}}

\def\gsim{{~\raise.15em\hbox{$>$}\kern-.85em
          \lower.35em\hbox{$\sim$}~}\xspace}
\def\lsim{{~\raise.15em\hbox{$<$}\kern-.85em
          \lower.35em\hbox{$\sim$}~}\xspace}

\def\pt         {\ensuremath{p_{\mathrm{T}}}\xspace}

\def\photos     {\mbox{\textsc{Photos}}\xspace}

\def\tell1  {TELL1\xspace}
\def\ukl1   {UKL1\xspace}

\newcommand{\eg}{\mbox{\itshape e.g.}\xspace}

\newcommand{\lhcborcid}[1]{\href{https://orcid.org/#1}{\hspace*{0.1em}\raisebox{-0.45ex}{\includegraphics[width=1em]{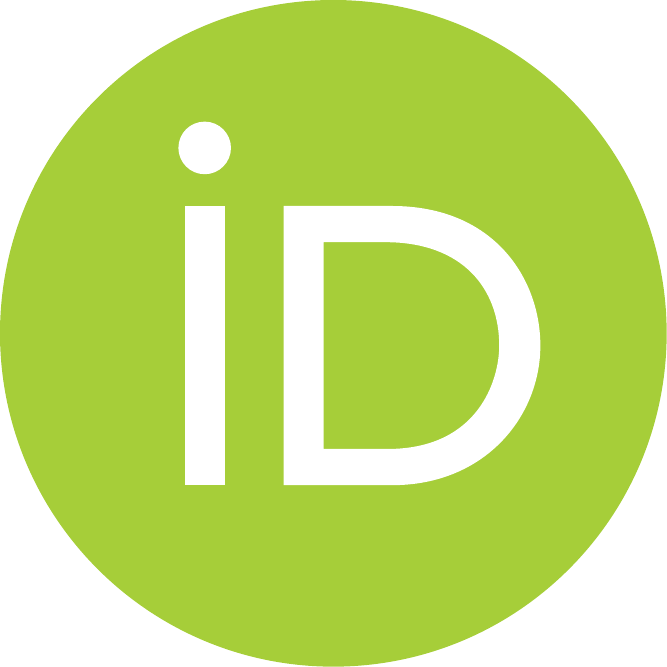}}}}

\usepackage{cite} 
\usepackage{mciteplus}

\def\RK         {\ensuremath{R_K}\xspace}
\def\RKst       {\ensuremath{R_{\Kstar}}\xspace}

\def\NRKorKst       {\ensuremath{R_{(K,\Kstar)}}\xspace}

\def\RJPsK     {\ensuremath{r_{\jpsi}^{\kaon}}\xspace}
\def\RJPsKst   {\ensuremath{r_{\jpsi}^{\Kstar}}\xspace}

\def\RPsiK{\ensuremath{R_{\psitwos}^{\kaon}}\xspace}
\def\RPsiKst{\ensuremath{R_{\psitwos}^{\Kstar}}\xspace}

\def\lqsq{low-\qsq\xspace}
\def\cqsq{central-\qsq\xspace}

\def\ll{\ensuremath{\ellp\ellm}\xspace}

\def\pp{\ensuremath{\proton\proton}\xspace}

\def\hepml {\mbox{hep\_ml}\xspace}

\def\wpid{\ensuremath{w_{\textrm{PID}}}\xspace}
\def\wtrk{\ensuremath{w_{\textrm{TRK}}}\xspace}

\def\wmco{\ensuremath{w_{\textrm{Mult\&Kin}}}\xspace}

\def\wlo{\ensuremath{w_{\textrm{\lone}}}\xspace}
\def\whlt{\ensuremath{w_{\textrm{\hlt}}}\xspace}
\def\wmcreco{\ensuremath{w_{\textrm{Reco}}}\xspace}

\def\BToKll{\decay{\B^{(+,0)}}{\kaon^{(+,*0)}\ll}}
\def\BToKee{\decay{\B^{(+,0)}}{\kaon^{(+,*0)}\epem}}
\def\BToKmm{\decay{\B^{(+,0)}}{\kaon^{(+,*0)}\mumu}}

\def\NBToKJPsll{\decay{\B^{(+,0)}}{\kaon^{(+,*0)}\jpsi(\decay{}{\ellell})}}

\def\NBToKPsill{\decay{\B^{(+,0)}}{\kaon^{(+,*0)}\psitwos(\decay{}{\ellell})}}

\def\NBToKJPsee{\decay{\B^{(+,0)}}{\kaon^{(+,*0)}\jpsi(\decay{}{\epem})}}

\def\NBToKJPsmm{\decay{\B^{(+,0)}}{\kaon^{(+,*0)}\jpsi(\decay{}{\mumu})}}

\def\BdToKstll{\decay{\Bd}{\Kstarz \ll}}

\def\BdToKstee{\decay{\Bd}{\Kstarz \epem}}

\def\BuToKll{\decay{\Bu}{\Kp \ll}}

\def\BuToKee{\decay{\Bu}{\Kp \epem}}

\def\lqsq{\ensuremath{\text{low-}\qsq}\xspace}
\def\cqsq{\ensuremath{\text{central-}\qsq}\xspace}

\def\wsmear{\ensuremath{w_{\textrm{Res}}}\xspace}

\usepackage{xfrac}
\usepackage{enumitem}
\usepackage{cleveref}
\usepackage{multirow}
\usepackage{placeins}
\usepackage{tikz}
\usetikzlibrary{trees,arrows,automata,shapes}
\usetikzlibrary{decorations.pathmorphing}
\usetikzlibrary{decorations.markings}
\usetikzlibrary{shapes.symbols}
\usetikzlibrary{shapes.geometric}
\usetikzlibrary{shapes.arrows}
\usetikzlibrary{positioning}
\tikzset
{
   vector/.style = {decorate, decoration={snake,amplitude=2pt, segment length=5pt}},
   fermion/.style = {postaction={decorate}, decoration={markings,mark=at position .55 with {\arrow{>}}}},
   fermionbar/.style = {postaction={decorate}, decoration={markings,mark=at position .55 with {\arrow{<}}}},
   gluon/.style = {decorate, decoration={coil,amplitude=3pt, segment length=5pt}},
   scalar/.style = {dashed, postaction={decorate}, decoration={markings,mark=at position .55 with {\arrow{>}}}},
   scalarbar/.style = {dashed, postaction={decorate}, decoration={markings,mark=at position .55 with {\arrow{<}}}}
}
\tikzstyle{information text}=[draw,rounded corners,inner sep=1ex]
\usepackage{longtable} 

\begin{document}

\renewcommand{\thefootnote}{\fnsymbol{footnote}}
\setcounter{footnote}{1}


\begin{titlepage}
\pagenumbering{roman}

\vspace*{-1.5cm}
\centerline{\large EUROPEAN ORGANIZATION FOR NUCLEAR RESEARCH (CERN)}
\vspace*{1.5cm}
\noindent
\begin{tabular*}{\linewidth}{lc@{\extracolsep{\fill}}r@{\extracolsep{0pt}}}
\ifthenelse{\boolean{pdflatex}}
{\vspace*{-1.5cm}\mbox{\!\!\!\includegraphics[width=.14\textwidth]{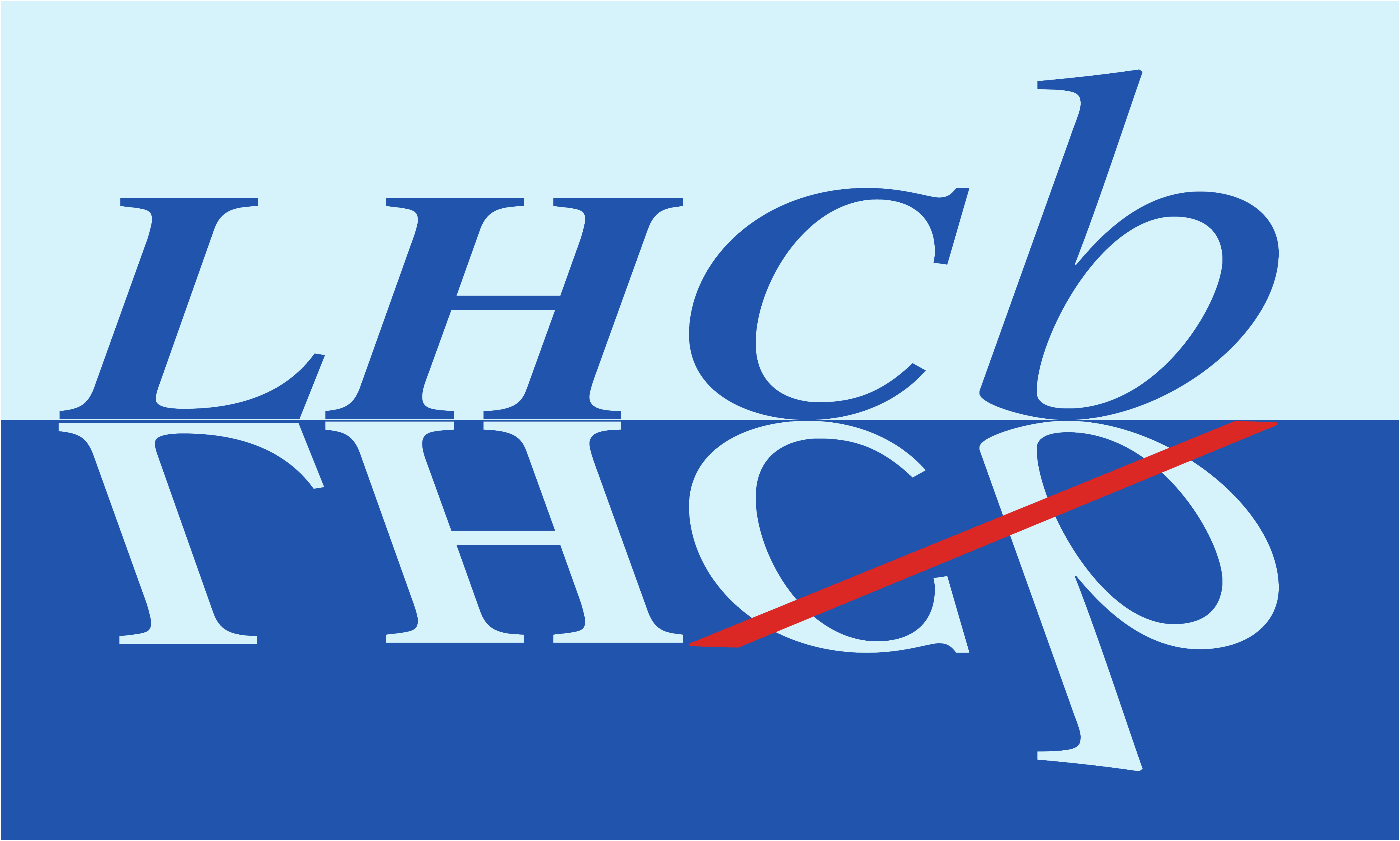}} & &}%
{\vspace*{-1.2cm}\mbox{\!\!\!\includegraphics[width=.12\textwidth]{figs/lhcb-logo.eps}} & &}%
\\
 & & CERN-EP-2022-277 \\  
 & & LHCb-PAPER-2022-046 \\  
 & & \today \\ 
 & & \\
\end{tabular*}

\vspace*{4.0cm}

{\normalfont\bfseries\boldmath\huge
\begin{center}
  \papertitle
\end{center}
}

\vspace*{1.0cm}

\begin{center}
\paperauthors\footnote{Authors are listed at the end of this paper.}
\end{center}

\vspace{\fill}

\begin{abstract}
  \noindent
  The first simultaneous test of muon-electron universality using \mbox{\BuToKll} and \mbox{\BdToKstll} decays is performed, in two ranges of the dilepton invariant-mass squared, \qsq. The analysis uses beauty mesons produced in proton-proton collisions collected with the \lhcb detector between 2011 and 2018, corresponding to an integrated luminosity of 9\invfb. 
  Each of the four lepton universality measurements reported is either the first in the given \qsq interval or supersedes previous \lhcb measurements. The results are compatible with the predictions of the Standard Model.
\end{abstract}

\vspace*{2.0cm}

\begin{center}
  Published in Phys. Rev. Lett. 131 (2023) 051803
\end{center}

\vspace{\fill}

{\footnotesize 
\centerline{\copyright~\papercopyright. \href{\paperlicenceurl}{\paperlicence}.}}
\vspace*{2mm}

\end{titlepage}


\newpage
\setcounter{page}{2}
\mbox{~}

\renewcommand{\thefootnote}{\arabic{footnote}}
\setcounter{footnote}{0}

\cleardoublepage


\pagestyle{plain} 
\setcounter{page}{1}
\pagenumbering{arabic}


Within the Standard Model of particle physics (SM), Lepton Universality (LU) states that the electroweak gauge bosons couple equally to all three families of leptons. Although LU is not protected by a fundamental conservation law in the SM, it has been validated experimentally at the percent and per mille levels in the decays of \W and \Z bosons, respectively~\cite{UA1:1988rck,CDF:1991mse,UA2:1992udr, CDF:1992jvc,D0:1995gzy,D0:1999bqi,ALEPH:2005ab,ALEPH:2013dgf,ATLAS:2016nqi,ATLAS:2020xea,CMS:2022mhs,LHCb-PAPER-2016-024,LHCb-PAPER-2018-016}. New particles in extensions of the SM may violate this symmetry, resulting in observable changes for the rates of quark-lepton transitions in the decays of SM hadrons.

In this context, rare,``nonresonant''  semileptonic \bsll decays, where $\ell$ represents either an electron or a muon, are  particularly interesting~\cite{Wang:2003je,Hiller:2003js}. 
The contribution of different operators to the SM decay rate varies as a function of the square of the dilepton invariant mass, \qsq. In the SM, differences in the decay rates to electron and muon final states originate  solely from lepton mass effects and are below the per mille level across the vast majority of final states and \qsq regions.
Modeling of final state radiation is the dominant source of uncertainty in the SM prediction of the observable difference between the electron and muon decay rates in \bsll processes, and there is consensus that this uncertainty is at most 1\%~\cite{Bordone:2016gaq,Isidori:2020acz,Isidori:2022bzw}.
In contrast, models extending the SM can introduce violation of LU in these transitions by more than 10\% while 
remaining compatible with all other experimental constraints~\cite{Hiller:2014yaa,Gripaios:2014tna,deMedeirosVarzielas:2015yxm,Barbieri:2016las,Altmannshofer:2014cfa,Crivellin:2015mga,Celis:2015ara,Falkowski:2015zwa}. Hence, any observed violation of LU 
 in \bsll decays would be an unambiguous sign of physics beyond the SM.
 Using current experimental data, these decays are sensitive to the existence of  particles beyond the SM at energy scales up to $\mathcal{O}(50\tev)$~\cite{LHCb-PII-Physics}.

In recent years there has been an accumulation of LU  measurements in \bsll transitions~\cite{LHCb-PAPER-2014-024,LHCb-PAPER-2017-013,LHCb-PAPER-2019-009,LHCb-PAPER-2019-040,LHCb-PAPER-2021-004,LHCb-PAPER-2021-038,Belle:2016fev,BaBar:2012mrf,BELLE:2019xld,Belle:2019oag} showing
a coherent pattern of deviations from the predictions of the SM. 
Notably, the most recent \lhcb LU test using \BuToKll 
decays~\cite{LHCb-PAPER-2021-004} reported evidence for violation of LU with a significance of 3.1$\sigma$ and a combined statistical and systematic uncertainty of around~5\%.
In addition, measurements of angular observables and decay rates in $\decay{b}{s\mumu}$ decays have also shown a similarly  intriguing pattern of deviations from SM predictions~\cite{LHCb-PAPER-2014-006,LHCb-PAPER-2015-009,LHCb-PAPER-2015-023,LHCb-PAPER-2016-012, LHCb-PAPER-2021-014,LHCb-PAPER-2021-022,LHCb-PAPER-2015-051,LHCb-PAPER-2020-002,LHCb-PAPER-2020-041,ATLAS:2018gqc,CMS:2015bcy,CMS:2017rzx}. 
A wide variety of phenomenological models
has been proposed to explain these measurements, \eg Refs.~\cite{Hiller:2014yaa,Gripaios:2014tna,deMedeirosVarzielas:2015yxm,Barbieri:2016las,Altmannshofer:2014cfa,Crivellin:2015mga,Celis:2015ara,Falkowski:2015zwa}, with many predicting the existence of observable \tev-scale particles.
It is therefore imperative to obtain a full understanding of the possible size of LU effects in \bsll transitions.

This Letter reports the first simultaneous  test of muon-electron universality in nonresonant 
\mbox{\BuToKll} and \mbox{\BdToKstll} decays that adopts a coherent approach to experimental effects, including 
cross-feed between different channels. 
A more comprehensive description of this test is reported in a companion article~\cite{LHCb-PAPER-2022-045}.
The inclusion of charge-conjugated processes is implied throughout the Letter. 
The $\Kstar(892)^0$ meson, denoted hereafter by \Kstarz, is reconstructed in the $\Kp\pim$ final state within 100\mevcc of its known  mass \cite{PDG2022}.  
The observables \RK and \RKst are defined in terms of the decay rates $\Gamma$ as
\begin{align}
   \NRKorKst(q^2_a,q^2_b) \, = &
     \,\, \frac
         {\displaystyle\int _{q^2_a}^{q^2_b}
            \frac{\deriv\Gamma(\BToKmm)}{\deriv\qsq}\deriv\qsq}
         {\displaystyle\int _{q^2_a}^{q^2_b}
            \frac{\deriv\Gamma(\BToKee)}{\deriv\qsq}\deriv\qsq } \;\;\;,
\end{align}
and measured in two \qsq intervals: $0.1<\qsq<1.1$\gevgevcccc, referred to as \lqsq, and $1.1<\qsq<6.0$\gevgevcccc referred to as \cqsq. 
All $\pp$ collision data recorded using the \lhcb detector between 2011 and 2018 are used, corresponding to integrated 
luminosities of 1.0, 2.0, and $6.0\invfb$ at center-of-mass energies of 7, 8, and $13\tev$, respectively.
In addition to the nonresonant decays in the \lqsq and \cqsq  intervals, 
``resonant'' \NBToKJPsll and \NBToKPsill decays are selected by requiring $6<\qsq<11$\gevgevcccc and $11<\qsq<15$\gevgevcccc, 
respectively, for electrons and by taking $\sqrt{\qsq}$  within  100\mevcc of the known \jpsi or \psitwos \cite{PDG2022} mass for muons. These resonant decays
are used both to calibrate the analysis and to reduce biases induced by differences in the detector response for electrons and muons.

Selection procedures for the 
\Kp\epem, \Kp\mumu, \Kp\pim\epem and \Kp\pim\mumu final states of \Bd and \Bu decays are harmonised to make the calibration of efficiencies as robust as possible.  This maximises the cancellation of efficiencies between the four processes in the final LU observables which, following previous analyses, are measured as double ratios,
\begin{align}
\NRKorKst & \equiv 
\frac{\tfrac{\cal{N}}{\varepsilon}(\BToKmm)}
     {\tfrac{\cal{N}}{\varepsilon}(\NBToKJPsmm)}
\bigg{/} 
\frac{\tfrac{\cal{N}}{\varepsilon}(\BToKee)}
     {\tfrac{\cal{N}}{\varepsilon}(\NBToKJPsee)} \, ,
\label{Eq:Rkorkst}
\end{align}
where $\tfrac{\cal{N}}{\varepsilon}(X)$ represents the efficiency corrected yield for process $X$.
These detector efficiencies are calibrated using resonant \NBToKJPsll decays;  the calibrations derived using \Kp and
\Kstarz final states are shown to be interchangeable for the first time. The  measurements of \RK and \RKst are optimized for statistical precision.  This leads to stringent particle identification (PID) requirements and dedicated multivariate selections to suppress backgrounds from hadronic and partially reconstructed \bquark-hadron decays, and a higher purity than previous \lhcb analyses of these final states~\cite{LHCb-PAPER-2021-004,LHCb-PAPER-2017-013}. 
Finally, data are used for the first time to measure residual backgrounds from misidentified \bquark-hadron decays for the final measurement of the \NRKorKst observables. These choices improve the statistical sensitivity per unit integrated luminosity and allow systematic uncertainties to be estimated more reliably than in previous \lhcb analyses of the same final states.

The \lhcb detector~\cite{LHCb-DP-2008-001,LHCb-DP-2014-002} is a single-arm forward
spectrometer covering the \mbox{pseudorapidity} range $2<\eta <5$. 
Simulated events produced with the software described in Refs.~\cite{Sjostrand:2007gs,*Sjostrand:2006za,LHCb-PROC-2010-056,Lange:2001uf} 
are used to model the effects of the detector acceptance, resolution, and the imposed selection requirements.
Dedicated simulated samples are produced for each year of data taking corresponding to the 
relevant detector and accelerator conditions. Final-state radiation is generated using \photos~\cite{davidson2015photos}. 
Samples of charm hadron, charmonia, and beauty hadron decays are collected 
for all data-taking periods and are used to calibrate the simulated single-particle reconstruction and PID efficiencies, ensuring that they match the performance of the detector. 
The real-time selection of \lhc bunch crossings (events) is performed by a trigger~\cite{LHCb-DP-2012-004}, 
which consists of a hardware stage, based on information from the calorimeter and muon
systems, followed by a software stage, which applies a full event
reconstruction. At the hardware trigger stage, events with large detector occupancy are rejected
and the remaining events are required to have a muon with high transverse momentum relative to the beamline (\pt), or a hadron or an electron with high 
transverse energy in the calorimeters. The software trigger requires a two- or three-body secondary vertex
with significant displacement from any primary  \pp interaction vertex. At least one charged particle
must have significant \pt and be inconsistent with originating from a primary vertex.
A multivariate algorithm~\cite{BBDT,LHCb-PROC-2015-018} based on kinematic, geometric and lepton identification criteria 
is used for the identification of secondary vertices consistent with the decay of a \bquark hadron.

Signal candidates are reconstructed by combining two  oppositely charged electrons or muons with either a \Kp or \Kstarz meson candidate.
Calorimeter clusters are identified as	bremsstrahlung associated with electron	candidates
if they	are consistent with a straight-line extrapolation of the electron candidate's trajectory
from the bremsstrahlung	emission point.

All final-state particles are required 
to be within the geometric acceptance of the PID  systems (\rich, calorimeters and muon detectors \cite{LHCb-DP-2014-002}) and the momentum range covered
by the PID calibration samples. They are also required to satisfy minimum \pt requirements and to have significant displacement from the primary \pp collision associated
with the signal candidate. A combination of requirements on individual PID  systems and 
multivariate classifiers, which combine information from all PID systems, is 
used to identify the final state particles. Identical PID requirements are used in all \qsq ranges. 

Two groups of multivariate classifiers, designed to be as similar as possible in their choice of features between the \Bu and \Bd signal decays, 
 are used to separate signal from residual backgrounds. 
The first group of classifiers is trained to separate simulated signal from random combinatorial backgrounds using 
kinematic and geometric features of the final-state particles and the signal candidate vertex, as well
as the vertex fit quality. Candidates passing all other selection requirements and having an invariant mass 
greater than 5400 (5600)\mevcc are used as a proxy for background when training the first muon (electron) classifiers.
A $k$-folding approach~\cite{Blum1999Jul} is used to avoid bias. The second group of classifiers
is trained to separate signal from the partially reconstructed \bquark-hadron decays \mbox{$\B\to (K,\Kstarz) \ll X$}, where $X$ represents one or more additional hadrons. Kinematic and geometric
features of the signal candidate are used, as well as features describing the spatial and kinematic isolation of 
the final-state particles and the signal candidate from other reconstructed particles.
Although a limited number of simulated backgrounds with a single missing pion are used for training, the classifiers have significant discrimination against most partially reconstructed decays. 

The working points of the two groups of classifiers are optimized simultaneously, using the expected signal significance for 
the predicted SM decay rates as the figure of merit. The optimal working point is chosen separately for each signal mode, \qsq region, lepton flavor, and data-taking period. For the electron mode, the ratio of the hadronic and 
dielectron momentum components transverse to the \B-meson  flight direction are used to correct the momentum 
of the dielectron pair~\cite{LHCb-PAPER-2017-013}. The resulting corrected mass has significant power
to separate signal from partially reconstructed backgrounds and is therefore used  in
the analysis, with the working point chosen analogously to the process followed for the multivariate classifiers.
As in previous \lhcb LU tests~\cite{LHCb-PAPER-2021-004,LHCb-PAPER-2017-013}, dedicated simulated event samples are used to study backgrounds 
that remain after all selection criteria have been applied. Specific vetoes are used to eliminate  
backgrounds resulting from misidentified \decay{\B}{\D(\decay{}{\kaon\pi})\electron\neue} and \decay{\B}{D(\decay{}{\kaon\electron\neue})\pi} decays.
Resonant \NBToKJPsll and \NBToKPsill decays in which a hadron and lepton of the same charge are swapped are
similarly vetoed. Finally, electron candidates are divided into three categories according to the number of bremsstrahlung photons (0, 1, $>1$) associated with the final-state \epem pair. 
After the application of all criteria, there are no nonresonant signal candidates in common between the \Bd and \Bu samples.
Around one percent of selected events have multiple candidates; in such cases a single reconstructed candidate is 
chosen randomly.

The \lhcb detector simulation is calibrated  to ensure an accurate description of differences in detector
response for electrons and muons. 
This is performed using an iterative multi-step process, where the output of 
each step $a$ is a set of weights $w_a$  characterizing the relative efficiency of data and simulation as a function of the 
signal candidate kinematics. Several of these steps use the same \NBToKJPsll resonant decays
that are subsequently  used in the analysis itself. The  weights are therefore computed for both \Bd 
and \Bu decays and  shown to be interchangeable.
To minimize correlations 
between the calibration samples and the candidates used in the final analysis,  weights derived from \Bd samples 
are used to correct \Bu efficiencies and vice versa.

Data calibration samples are used to evaluate PID  efficiencies for hadrons, muons, and electrons 
as a function of particle momentum, pseudorapidity and the detector occupancy. Electron identification efficiencies 
are also evaluated separately for candidates having one or no associated bremsstrahlung photons. Simulation
is used to verify that PID requirements factorize, so that the derived data-simulation weights \wpid 
can be applied independently of one another. Hadron and muon particle identification efficiencies are found to 
factorize, whereas for electrons with overlapping calorimeter clusters the efficiencies depend strongly on both \qsq and the separation of the
electron trajectories extrapolated to the calorimeter. The electron and positron candidates are required to be well separated on entering the calorimeters.
Particle reconstruction efficiencies are found to factorize for 
all particle species. Muon and hadron reconstruction efficiencies are found to be described well by simulation and do not require calibration, while electron
reconstruction efficiencies are calibrated using  weights \wtrk in intervals of momentum, pseudorapidity and in two geometric regions of the
vertex detector that encode the amount of material traversed by the electron. 

The kinematics of the \B meson and the multiplicity of the underlying event are calibrated using weights \wmco by training a
boosted decision tree (BDT) \cite{Rogozhnikov:2016bdp} with the \texttt{\hepml} library \cite{alex_rogozhnikov_2021_5500612}. The event track multiplicity and the signal candidate \pt, 
momentum and pseudorapidity are used
as inputs. The BDT is trained using \mbox{\NBToKJPsmm} candidates selected in data 
and using a specific trigger path that is fully aligned between data and simulation in all data-taking periods.  The hardware trigger efficiencies are calibrated using weights \wlo  
as a function of muon \pt and electron transverse energy, while the software trigger
efficiencies are calibrated using weights \whlt   as a function of event track multiplicity. Finally, 
 residual data-simulation differences are corrected by weights \wmcreco from a second BDT,  trained  using the same kinematic variables
as the first BDT and the \chisqip of the \B and \jpsi mesons, with \chisqip defined as the  difference in \chisq of the primary \pp  collision vertex with and without the considered particle.

The invariant-mass resolution of the simulated electron data must also be calibrated using weights \wsmear 
so that the migration of candidates between \qsq 
regions can be evaluated accurately. This calibration is derived by fitting the dielectron invariant mass 
component of the fully selected resonant \NBToKJPsee candidates in data and simulation. A modified Crystal Ball 
function~\cite{Skwarnicki:1986xj} with power-law tails both above and below the mean mass value is used to model 
the dielectron spectrum, with  the remaining combinatorial background modeled using an exponential function.

\begin{figure}[ht]
    \centering
    \includegraphics[width=0.99\linewidth]{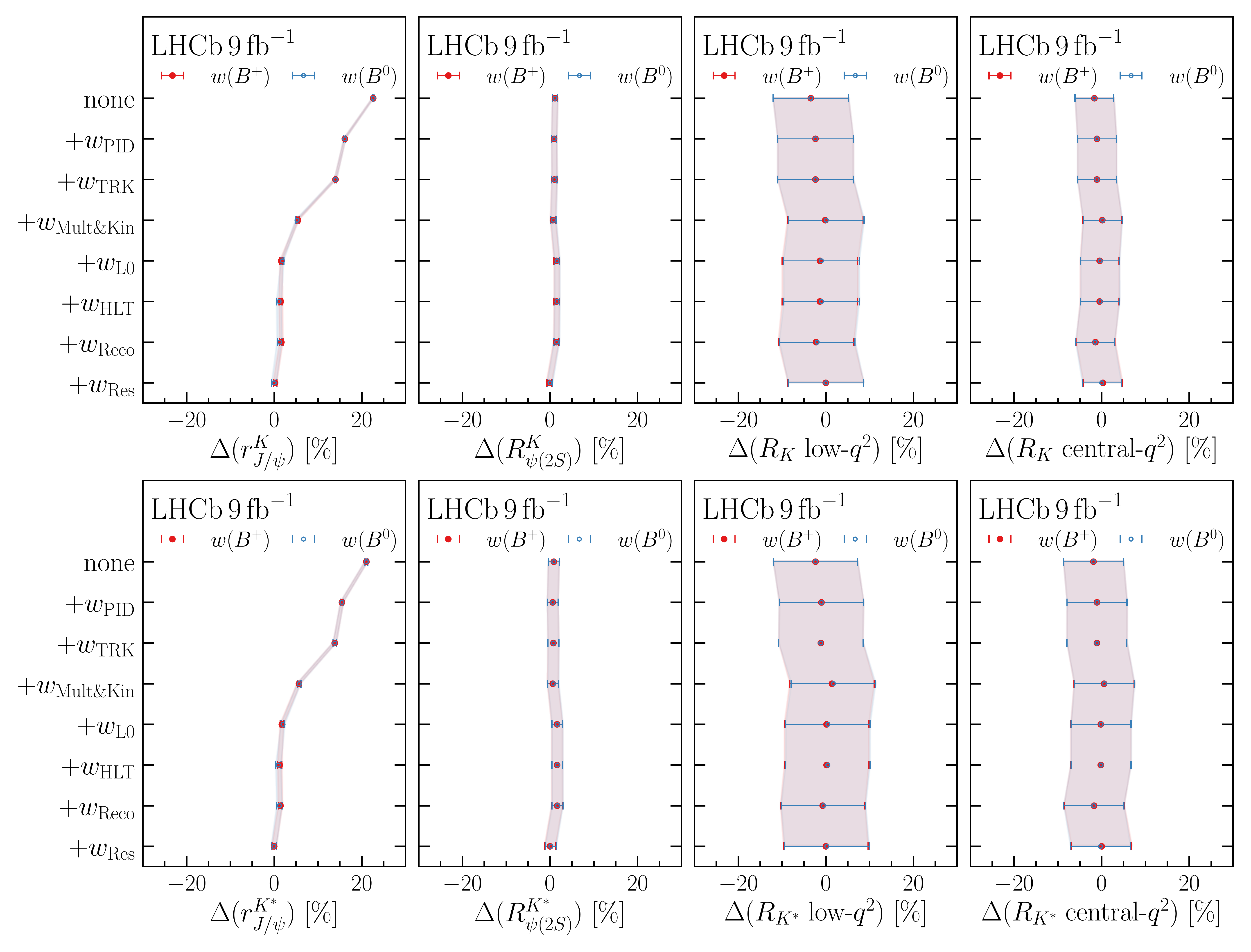}
    \caption{Variation of efficiency corrected ratios (top row, left to right): \RJPsK, \RPsiK, \RK in low- and \cqsq regions; (bottom row) analogous plots for the \Kstarz channel. Measurements are relative to the final value for the given observable. The simulation calibration using \Bd and \Bu chains for successive calibration weights are shown and almost indistinguishable.}
    \label{fig:single_double_ratio_stability}
\end{figure}

Four efficiency corrected resonant mode yield ratios are defined as benchmarks for this analysis: \RJPsK and \RPsiK using \Bu candidates;  \RJPsKst and \RPsiKst using \Bd candidates.
The single ratios \RJPsK and \RJPsKst represent the yields of \NBToKJPsmm decays relative to \mbox{\NBToKJPsee} decays. The double ratios \RPsiK and \RPsiKst are analogous to Eq.~\ref{Eq:Rkorkst} but with \NBToKPsill in place of the nonresonant \BToKll decays.
Resonant mode yields are extracted using a simultaneous maximum-likelihood fit to the invariant mass of the \B candidate. 
Signal lineshapes, as well as misidentified and partially reconstructed background shapes, are derived using simulated data.
Residual data-simulation differences are parametrized by a shift in the mean and a scale factor applied to the width of the signal lineshape. Remaining combinatorial backgrounds are modeled using an exponential function with a freely varying slope.
Ratios of misidentified background yields are constrained between the electron and muon final states using efficiencies derived from simulated data. The measured single ratios are found to be $\RJPsK = 1.047 \pm 0.024$ and $\RJPsKst = 1.028 \pm 0.024$, 
while the measured double ratios are found to be $\RPsiK = 0.987 \pm 0.007$ and $\RPsiKst = 1.012 \pm 0.013$. 
These uncertainties include
both statistical and systematic components: the latter dominate for the single ratios, while the opposite is true for the double ratios. 
The \RJPsK and \RJPsKst ratios have no significant dependence on kinematic and geometrical quantities, and a systematic uncertainty is assigned for residual non-flatness. 
As in previous \lhcb lepton universality tests, agreement of these quantities with 
predictions of the SM was a prerequisite to evaluating the \NRKorKst observables.

Figure~\ref{fig:single_double_ratio_stability} illustrates the 
evolution of the efficiency corrected ratios
 \RJPsK, \RJPsKst, \RPsiK and \RPsiKst 
as a function of the 
corrections applied to the simulated data. Since measurements of these resonant decays are consistent with LU \cite{PDG2022}, 
they are excellent control modes for this analysis. Although the \Bd and \Bu weights are found to be interchangeable,
the 25\% effect observed in the efficiency corrected single ratios leads to undesirably large systematic uncertainties. These are reduced by defining the final observables using the  double ratio of Eq.~\ref{Eq:Rkorkst}.

Figure~\ref{fig:single_double_ratio_stability} also shows the evolution of this double ratio for each of the two observables and \qsq ranges.
Contrary to the single ratio, the impact of the calibration chain on the double ratio is less than a few percent,
significantly reducing systematic uncertainties associated with the determination of efficiencies from 
simulated data. 
The double ratio is evaluated using a simultaneous maximum-likelihood fit to the 
\Bu and \Bd candidate invariant masses in the \lqsq, \cqsq and \jpsi regions, 
where the efficiencies are constrained using values 
and uncertainties obtained from simulated data.

After all selection criteria have been applied, the nonresonant muon candidate samples contain only signal and combinatorial backgrounds.
The invariant mass spectrum of the nonresonant muon signals is modeled using simulated data, while the combinatorial
background is modeled using an exponential function with a freely varying slope. Small adjustments to the mean and width of the  
signal distribution are constrained to be the same as for the resonant muon signal. The normalization of both components is allowed to vary freely.

The nonresonant electron candidate samples have a more complex composition. In addition to signal and combinatorial background they 
contain resonant \jpsi meson decays that migrate into the \cqsq (but not \lqsq) region (``leakage''), partially reconstructed decays without 
misidentification, and residual misidentified hadronic decays. The invariant mass spectra of the nonresonant electron signals are modeled following a similar  procedure to the muon modes and over a wider mass range than previous analyses, reducing correlations between components. The combinatorial background is modeled using an exponential function with 
freely varying normalization multiplied by a function that parametrizes the distortion by the 
corrected mass 
and partially reconstructed selection criteria. 
The parameters of this function are fixed from background data reconstructed using two leptons of the same charge and varied in pseudoexperiments to attribute an associated systematic uncertainty.
The leakage is modeled using a 
kernel-density estimator \cite{Cranmer:2000du} derived from simulated data; its normalization is constrained to the observed 
resonant mode yields multiplied by efficiencies measured in simulation. The partially reconstructed backgrounds are 
modeled using kernel-density estimators derived from simulated data, with a systematic uncertainty assigned for 
the finite number of different partially reconstructed decays that are simulated. Their normalization is allowed to vary freely for the 
\Bd nonresonant candidate samples. In the \Bu samples, the normalization is constrained to the observed nonresonant \Bd signal yields in the 
same \qsq region multiplied by efficiencies measured in simulation and accounting for isospin partner modes. 

\begin{figure}[t]
    \centering
    \includegraphics[width=0.95\linewidth]{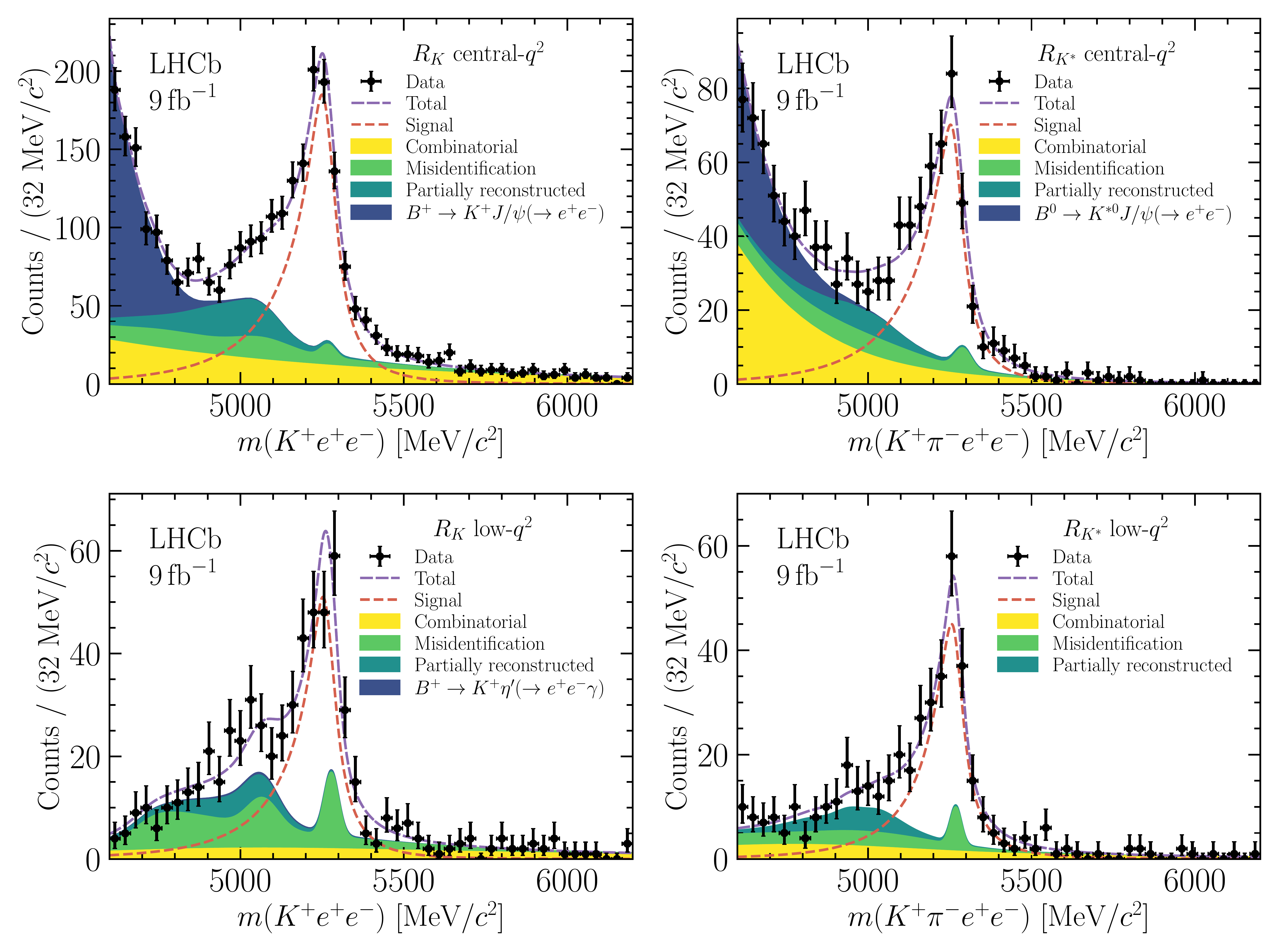}
    \caption{Invariant mass distributions of the nonresonant (left) \mbox{\BuToKee} and (right)  \mbox{\BdToKstee} candidates in (upper) \cqsq and (lower) \lqsq regions. The results of the fit described in the text are also presented.} %
    \label{fig:electron_rare_mass_shapes}
\end{figure}

Residual misidentified backgrounds may contain one or two hadrons misidentified as electrons, as well as additional missing energy.
A wide range of these backgrounds are studied using simulated data and each individual background is found to be small
compared to the expected statistical sensitivity. However, the large number of potential hadronic \B decays, which in many cases 
have poorly known or unknown dynamics, means that their overall contribution to the final candidate sample is not necessarily small. 
Data are therefore used to estimate the overall invariant mass spectra and normalization of such backgrounds.
A sample enriched with background from misidentification of nonresonant candidates is defined for each of the four LU  observables by inverting the 
electron identification criteria and using less stringent electron identification requirements. This sample is corrected for the
residual contribution of signal decays  and then weighted using  misidentification rates  measured from data to obtain
the expected misidentified backgrounds that would pass all selection criteria. The invariant mass shape of these backgrounds
is modeled using an empirical function; 
their normalization is constrained
using the measured central values and uncertainties.

The invariant mass distributions of the nonresonant electron candidates resulting from the final fit to the four LU  observables
are shown in Fig.~\ref{fig:electron_rare_mass_shapes}. The measured values of the observables of interest are
\begin{align*}
    \lqsq \begin{cases}
        \RK   &= 0.994~^{+0.090}_{-0.082} \stat ^{+0.029}_{-0.027} \syst, \\
        \RKst &= 0.927~^{+0.093}_{-0.087} \stat ^{+0.036}_{-0.035}\syst, \\
   \end{cases} \\
   \cqsq 
   \begin{cases}
        \RK  &= 0.949~^{+0.042}_{-0.041} \stat ^{+0.022}_{-0.022} \syst, \\
        \RKst &=1.027~^{+0.072}_{-0.068} \stat ^{+0.027}_{-0.026} \syst.
   \end{cases} 
\end{align*}
\noindent All four measurements are in agreement with predictions of the SM~\cite{Bordone:2016gaq,Isidori:2020acz,Bobeth:2007dw,Straub:2018kue,vanDyk:2021sup,Capdevila:2016ivx,Capdevila:2017ert,Serra:2016ivr,Altmannshofer:2017fio,Jager:2014rwa,Ghosh:2014awa}.

Systematic uncertainties associated with efficiencies are evaluated by varying the assumptions 
made when calibrating the simulated samples. The biggest uncertainty of this type is the stability of 
the \RJPsK and \RJPsKst ratios as a function of different kinematic and geometric variables associated with 
these decays. The overall systematic uncertainties for efficiencies are below 1\% in all cases except \RKst \lqsq where they are 2\%.
Systematic uncertainties associated with the modeling of nonresonant decay form factors, which affect the efficiencies, are evaluated using simulation 
and found to be negligible for \Bu decays and around 1\% for \Bd decays. Systematic uncertainties
associated with the modeling of the invariant mass distributions are dominated by the data-driven modeling
of misidentified backgrounds, and are \mbox{2.0\%--2.5\%} depending on the LU observable in question. 
Although larger than the systematic uncertainties assigned in the previous \lhcb analyses \cite{LHCb-PAPER-2021-004, LHCb-PAPER-2017-013}, these are significantly smaller than the statistical uncertainties associated with each of the four LU observables.

The results presented here differ from previous \lhcb measurements of \RK~\cite{LHCb-PAPER-2021-004} and \RKst~\cite{LHCb-PAPER-2017-013}.
In the case of \RKst the data sample is much larger compared to the previous measurement, and hence statistical effects can explain the majority of the difference.
For \RK \cqsq, on the other hand, the change is primarily due to systematic effects; 
the biggest shift (0.064) comes from the use of more stringent electron identification criteria which greatly reduce the contribution from processes that had previously not been accounted for appropriately, 
while a further shift (0.038) comes from the modeling of the residual misidentified background.

The measurements of \RK and \RKst for the 
 \qsq intervals $0.1<\qsq<1.1$\gevgevcccc and $1.1<\qsq<6.0$\gevgevcccc reported here supersede previous \lhcb measurements \cite{LHCb-PAPER-2021-004,LHCb-PAPER-2017-013} and are in agreement with the predictions of the SM.
The systematic uncertainties associated with these measurements remain significantly smaller than the statistical uncertainties and are expected to reduce further with more data.

\newpage
\section*{Acknowledgements}
%
%
\noindent We express our gratitude to our colleagues in the CERN
accelerator departments for the excellent performance of the LHC. We
thank the technical and administrative staff at the LHCb
institutes.
We acknowledge support from CERN and from the national agencies:
CAPES, CNPq, FAPERJ and FINEP (Brazil); 
MOST and NSFC (China); 
CNRS/IN2P3 (France); 
BMBF, DFG and MPG (Germany); 
INFN (Italy); 
NWO (Netherlands); 
MNiSW and NCN (Poland); 
MEN/IFA (Romania); 
MICINN (Spain); 
SNSF and SER (Switzerland); 
NASU (Ukraine); 
STFC (United Kingdom); 
DOE NP and NSF (USA).
We acknowledge the computing resources that are provided by CERN, IN2P3
(France), KIT and DESY (Germany), INFN (Italy), SURF (Netherlands),
PIC (Spain), GridPP (United Kingdom), 
CSCS (Switzerland), IFIN-HH (Romania), CBPF (Brazil),
Polish WLCG  (Poland) and NERSC (USA).
We are indebted to the communities behind the multiple open-source
software packages on which we depend.
Individual groups or members have received support from
ARC and ARDC (Australia);
Minciencias (Colombia);
AvH Foundation (Germany);
EPLANET, Marie Sk\l{}odowska-Curie Actions and ERC (European Union);
A*MIDEX, ANR, IPhU and Labex P2IO, and R\'{e}gion Auvergne-Rh\^{o}ne-Alpes (France);
Key Research Program of Frontier Sciences of CAS, CAS PIFI, CAS CCEPP, 
Fundamental Research Funds for the Central Universities, 
and Sci. \& Tech. Program of Guangzhou (China);
GVA, XuntaGal, GENCAT and Prog.~Atracci\'on Talento, CM (Spain);
SRC (Sweden);
the Leverhulme Trust, the Royal Society
 and UKRI (United Kingdom).

\bibliographystyle{LHCb}
\bibliography{main,standard,LHCb-PAPER,LHCb-CONF,LHCb-DP,LHCb-TDR}
\newpage
\centerline
{\large\bf LHCb collaboration}
\begin
{flushleft}
\small
R.~Aaij$^{32}$\lhcborcid{0000-0003-0533-1952},
A.S.W.~Abdelmotteleb$^{50}$\lhcborcid{0000-0001-7905-0542},
C.~Abellan~Beteta$^{44}$,
F.~Abudin{\'e}n$^{50}$\lhcborcid{0000-0002-6737-3528},
T.~Ackernley$^{54}$\lhcborcid{0000-0002-5951-3498},
B.~Adeva$^{40}$\lhcborcid{0000-0001-9756-3712},
M.~Adinolfi$^{48}$\lhcborcid{0000-0002-1326-1264},
P.~Adlarson$^{77}$\lhcborcid{0000-0001-6280-3851},
H.~Afsharnia$^{9}$,
C.~Agapopoulou$^{13}$\lhcborcid{0000-0002-2368-0147}
% ...

{\footnotesize \it

% ...
$^{9}$Universit{\'e} Clermont Auvergne, CNRS/IN2P3, LPC, Clermont-Ferrand, France\\
% ...
$^{13}$LPNHE, Sorbonne Universit{\'e}, Paris Diderot Sorbonne Paris Cit{\'e}, CNRS/IN2P3, Paris, France\\
% ...
$^{32}$Nikhef National Institute for Subatomic Physics, Amsterdam, Netherlands\\
% ...
$^{40}$Instituto Galego de F{\'\i}sica de Altas Enerx{\'\i}as (IGFAE), Universidade de Santiago de Compostela, Santiago de Compostela, Spain\\
% ...
$^{44}$Physik-Institut, Universit{\"a}t Z{\"u}rich, Z{\"u}rich, Switzerland\\
% ...
$^{48}$H.H. Wills Physics Laboratory, University of Bristol, Bristol, United Kingdom\\
% ...
$^{50}$Department of Physics, University of Warwick, Coventry, United Kingdom\\
% ...
$^{53}$School of Physics and Astronomy, University of Glasgow, Glasgow, United Kingdom\\
$^{54}$Oliver Lodge Laboratory, University of Liverpool, Liverpool, United Kingdom\\
% ...
$^{77}$Department of Physics and Astronomy, Uppsala University, Uppsala, Sweden, associated to $^{53}$\\
% ...
}
\end{flushleft}
\end{document}